%% file: main.tex
\documentclass[fleqn,10pt]{SelfArx} % Document font size and equations flushed left

\usepackage[english]{babel} % Specify a different language here - english by default

\usepackage{natbib}
\usepackage[table]{xcolor}
\usepackage{wrapfig}
\newcommand{\arcsec}{$^{\prime\prime}$}

\definecolor{color1}{RGB}{0,0,90} % Color of the article title and sections
\definecolor{color2}{RGB}{0,20,20} % Color of the boxes behind the abstract and headings

\usepackage{hyperref} % Required for hyperlinks
\hypersetup{hidelinks,colorlinks,breaklinks=true,urlcolor=color2,citecolor=color1,linkcolor=color1,bookmarksopen=false,pdftitle={Title},pdfauthor={Author}}

\JournalInfo{White paper on High-Resolution X-ray Imaging} % Journal information
\PaperTitle{Star formation, stellar evolution, and planets in high-resolution X-ray imaging} % Article title
\Archive{}

\Authors{Hans Moritz G\"unther\textsuperscript{1}*, Scott J. Wolk\textsuperscript{2}, Sean J. Gunderson\textsuperscript{1}, Ruchi Pandey\textsuperscript{3,4}, Keivan G.\ Stassun\textsuperscript{5},  Lynne Valencic\textsuperscript{3,4}}% Authors
\affiliation{\textsuperscript{1}\textit{Massachusetts Institute of Technology}} % Author affiliation
\affiliation{\textsuperscript{2}\textit{Smithonian Astrophysical Observatory}} % Author affiliation
\affiliation{\textsuperscript{3}\textit{Johns Hopkins University}} % Author affiliation
\affiliation{\textsuperscript{4}\textit{NASA Goddard Space Flight Center}} % Author affiliation
\affiliation{\textsuperscript{5}\textit{Vanderbilt University}} % Author affiliation

\affiliation{*\textbf{Corresponding author}: hgunther@mit.edu} % Corresponding author

 % Keywords - if you don't want any simply remove all the text between the curly brackets
\newcommand{\keywordname}{Keywords} % Defines the keywords heading name

\Abstract{Stars set the conditions for planet formation, planet evolution, and habitability -- all major topics in astronomy today. Stars are also important in their own right as the most visible component of galaxies. In cool stars, X-ray emission is powered by magnetic fields, and so far our Sun is the only system in which those fields are spatially resolved. High-resolution X-ray (HiReX) imaging can track the origin and evolution of those fields, see how they connect young stars to their disks and outflows, and measure the energy, mass, and momentum that radiation and coronal mass ejections (CMEs) carry into the circumstellar environment. This is crucial for understanding whether planets can form and survive in young stellar systems, whether life can develop on those planets, and how the star evolves over time.

Intermediate-mass and high-mass stars blow winds and eventually evolve into degenerate objects such as white dwarfs, neutron stars, and black holes. Their evolution and death drive the chemical evolution of galaxies. High-resolution X-ray (HiReX) imaging can study the hottest components in those systems, such as the colliding winds of massive stars, accretion and nova explosions in CVs, and the shocks in outflows that form planetary nebulae. All these cases have in common that the X-ray emission is tracing the hottest, fastest, and most energetic components of the shocks. HiReX observations can reveal the temperature, spatial structures, and elemental abundances of different system components that no other wavelength can.

While stars are physically small compared to more powerful objects such as accreting black holes and AGN, they are also much closer to us, allowing a HiReX mission to resolve a variety of physical phenomena fundamental to our understanding of how stellar systems form, evolve, and interact with their environment.
}

\begin{document}

\flushbottom % Makes all text pages the same height

\maketitle % Print the title and abstract box
%\tableofcontents % Print the contents section

\thispagestyle{empty} % Removes page numbering from the first page

\section{Overview of science cases}
There are several ways to look at science cases for high-resolution X-ray (HiReX)  imaging. When designing an instrument, one has to ask ``What science can I do with a specific spatial resolution?'' but looking from a perspective of specific science areas the statement becomes ``Which questions do we need to answer to understand the universe?'' We provide a summary in Table~\ref {tab:science} that is roughly sorted by the required spatial resolution, but in the following sections we will analyze the open questions in star formation, stellar evolution, and planets that require HiReX imaging for each subfield.

% Add column with section numbers
% 

\begin{table*}[h!]
    \centering
    \rowcolors{1}{gray!10}{white}
    \begin{tabular}{{p{2.0cm}p{6.3cm}p{1.4cm}p{5cm}p{0.8cm}}}
        Topic & Science Case & Resolution & Other Requirements & section\\
        \hline
        Solar System & Resolve comets & 0.1$^{\prime\prime}$ & large $A_\mathrm{eff}$ & \ref{sect:comets}\\
        & Planetary Aurorae & 0.005$^{\prime\prime}$ & & \ref{sect:aurorae} \\
        Planetary Nebulae & Observe mixing in fast wind and nebula, radial distribution of hot bubble mission & 0.1$^{\prime\prime}$ & soft $A_\mathrm{eff} > 10\times$Chandra for faint features, FOV: arcmin, 0.2$-$2 keV & \ref{sect:PN}\\
        
        Star Formation & Proto-stellar jet launching & 0.01$^{\prime\prime}$ & 0.5$-$2.0 keV & \ref{sect:jets} \\
                       & distant young stellar clusters & 0.01$^{\prime\prime}$ & & \ref{sect:protostars}\\

        Novae & Resolve ejecta from super-soft source & 0.05$^{\prime\prime}$ & 0.1$-$2 keV, TOO: hours, then repeat visits over weeks/months & \\
            & Locate X-rays emission between WD and MS star & 0.01$^{\prime\prime}$ &  & \\

        %(Exo-)planets & Planetary aurorae & 0.001$^{\prime\prime}$ & 0.2$-$15 keV, large $A_\mathrm{eff}$\\
         (Exo-)planets   & Exoplanet plasmoid & 0.001$^{\prime\prime}$ & \ref{sect:CV} & \ref{sect:plasmoid}\\

        Activity & Astrospheres & 0.1$^{\prime\prime}$ & 0.2$-$2.0 keV & \ref{sect:astrosphere}\\
            & Resolving stellar coronae & $10^{-5\;\prime\prime}$ & large $A_\mathrm{eff}$ &\\
            & Coronal superflares & $10^{-5\;\prime\prime}$ & FOV: $>$arcmin$^2$ or TOO: hours, 2$-$7 keV & \ref{sect:flares}, \ref{sect:flaredisk} \\
            & Extreme coronal mass ejections (CMEs) & $10^{-5\;\prime\prime}$ &  & \ref{sect:CMEs}\\
            & Magnetically interacting binary stars & $<10^{-4\;\prime\prime}$ &\\

        Magnetic CVs & Resolve accretion column & $10^{-4\;\prime\prime}$ & & \ref{sect:CV}\\

        Massive Stars & Distribution of wind shocks & $10^{-4\;\prime\prime}$ & $0.2-2.0$\,keV & \ref{sect:massive}\\
            & Structure of colliding wind binary shock cones & $\leq10^{-4\;\prime\prime}$ & More distance limited; repeat over orbital period & \ref{sect:massive}\\
            & Topology of magnetic fields & $10^{-4\;\prime\prime}$ & large $A_\mathrm{eff}$ & \ref{sect:massive}\\

    \end{tabular}
    \caption{Summary of stellar science cases and driving requirements. Most science cases are discussed in detail in the following sections.}
    \label{tab:science}
\end{table*}

\section{Planets and Comets}
The only planetary system that we know for sure to be habitable to life is our own. It is thus critically important to study how energetic radiation interacts within our system to understand its impact on planets, atmospheres, and biomolecules in space.
Planets and comets in our own solar system emit X-rays via charge exchange with the solar wind. A HiReX mission can remotely resolve features that are currently limited to very rare and expensive in-situ measurements (e.g., Juno around Jupiter) to study space weather, atmospheres, and mass loss. The brightest of those phenomena are also observable in exoplanet systems with sufficient spatial resolution, where they open a window into processes that will never be accessible to in situ studies.

\subsection{Comets}
\label{sect:comets}
Comets are engines that convert Solar wind Ions to X-rays at discrete wavelengths via charge exchange with the neutral, moderate-density medium evaporating from the comet.
As such, they contain information about the solar wind at the comet's location and the state of the cometary coma. In the Chandra era, comets have been under-resolved, as we often require a 5\arcsec{} region to detect 5 photons, so the key here will be collecting area.  On the other hand, the payoff is a detailed map of the comet's coma.  Since it will be optical, we can trace features down to the comet's surface, including the possibility of identifying vents. To resolve a vent 100 meters across at a distance of 0.2 AU requires a resolution of
0.0003\arcsec.

\subsection{Jupiter Aurorae}
\label{sect:aurorae}
One can imagine observing Jovian or Saturnian aurora (and fluorescence). The auroral luminosity of Jupiter is about $10^{16}$ erg/s. X-ray fluorescence is similar. So the auroral flux is $\sim 2 \times 10^{-13}\;\mathrm{erg}/\mathrm{cm}^2/\mathrm{s}$.  The structure is large, covering 10,000 km in length, but narrow, with bead-like features no larger than 300 km.  At closest approach between the Earth and Jupiter, this is about 0.1\arcsec.  More commonly, these features require a resolution of 0.08\arcsec. The Jovian Auroral Distributions Experiment (JADE) on Juno has detected fluctuations occurring on timescales of a second or less. These rapid shifts imply that individual auroral filament structures are as narrow as a few tens of kilometers \citep{Allegrini_2017}. To resolve these from Earth requires a resolution of the order of 0.005\arcsec.
%JGR "Electron beams and loss cones in the auroral regions of Jupiter" F. Allegrini et al. 2017

\subsection{Exoplanet Plasmoid Rings}
\label{sect:plasmoid}
Atmospheric escape of close-in exoplanets driven by stellar irradiation influences their evolution, composition, and atmospheric dynamics.  There are several examples of atmospheric escape in the literature, starting quite nearby (Earth, Mars) and extending to spectacular examples including WASP 121 and HD~209458, both of which have had cometary tails detected via transit \citep{Gascon_2025}. These tails should primarily consist of low-density plasma, which should emit in soft X-rays, either due to superheating by the kinetic impact of the wind or via charge exchange. The plasmoids should form partial rings around the star.  In the case of HD~209458, at 48 pc and with a separation of 0.05 AU, the separation between the star and the ring is about 0.001\arcsec.

\section{Astrospheres}
\label{sect:astrosphere}
An astrosphere is the region of a stellar system occupied by the host star's stellar wind. It can be detected in X-rays as soft charge exchange (CX) lines \citep{2024NatAs...8..596K,2026ApJ...999..125L} and in the ultraviolet (UV) as excess emission in the hydrogen Lyman-$\alpha$ profile, as the highly ionized stellar winds interact with neutral gas \citep{2019ApJ...886...41L}.
Our Sun is obviously the best-studied system \citep{2012AN....333..324D}.
The winds of Sun-like main-sequence stars are relatively modest but have a tremendous impact on carving out a host-star-dominated bubble in the galactic ISM. Inside this bubble, surfaces and atmospheres of bodies are directly affected by the stellar wind,
not the ISM. (For example, stellar winds can erode planetary atmospheres through sputtering and ion pick-up, or redden, reduce, and erode airless surfaces through ``space weathering''.) Every star with an appreciable wind will have/support an astrosphere. The solar system's astrosphere, aka ``the heliosphere'', is ~120 au in radius, according to in situ measurements by Voyager spacecraft.

At a resolution of $0.1^{\prime\prime}$ we could easily resolve solar-like astrospheres to $>100$~pc and probe their shape, providing a wide range of targets with different ages, spectral types, activity levels, and planets. For close targets, we can constrain the thickness of the termination shock \citep{2026ApJ...999..125L}. That paper detected HD~61005 and estimates about 1.5~cts/s with Chandra/ACIS. The target is embedded in a region of dense ISM, which favors CX between the stellar wind and the ISM. For meaningful, spatially resolved astrosphere observations, a large effective area at 0.5-2.0~keV is required, where a Chandra-like $A_\mathrm{eff}$ allows the study of stars with strong winds in dense ISM regions and a larger $A_\mathrm{eff}$ enables the study of a wider range in target ages and wind mass loss rates.

\section{Magnetic fields in young or main-sequence cool stars}
Exoplanets are obviously the places where we look for life in the universe, and their host stars can make planets hospitable to life or destroy all hope of finding it. The formation of stars and planets is intricately linked with planets evolving in the disks that surround young stars. HiReX imaging can resolve hot, ionized regions in these systems, such as host-star coronae, and determine how they move and how they impact the evolution of stars, planets, and possibly life. On the other hand, in order to make a planet habitable, you need to form a star and planet in the first place, and a HiReX mission can uniquely answer outstanding questions in star formation. All of this directly addresses the science priority area ``Pathways to habitable worlds'' in the NASA Astrophysics Decadal Survey 2020 \citep{2021pdaa.book.....N}.

\subsection{Resolving stellar flares}
\label{sect:flares}
Magnetic reconnection powers solar and stellar flares. These flares emit X-rays, which can alter the chemistry \citep{2022A&A...667A..15K}, evaporate the atmospheres of close-in exoplanets \citep{2018MNRAS.479.5012O}, and destroy complex molecules, thereby preventing the evolution of life \citep{2019ApJ...881..114Y}. Stellar activity is also used as a proxy for stellar age to date the evolutionary status of stellar systems. Yet our Sun is the only star whose flares we can spatially resolve. We scale up those models by five orders of magnitude or more to interpret spectra and lightcurves from stars with vastly different activity levels, different magnetic structure, and different evolutionary history. Only high-resolution X-ray images can tell us if these hot flares actually behave as scaled-up versions of the Sun or if our interpretation of the geometry is fundamentally wrong. So far, we only have indirect information from line kinematics, and a few special cases where, e.g.\ eclipses or transits constrain the flare location or size \citep[e.g.][]{2003A&A...412..849S}.
Large flares are powerful events, so they are easily observable in nearby systems. A range of targets can be studied at different scales. The most easily accessible ones are Algol-like binaries where we expect flares connecting both stars (in Algol itself, the binary separation is 2~marcs, so we need spatial resolution of order 0.1~marcs) or giant stars which can have exceptionally long flare loops (e.g., the superflare on HD~251108 seen by \cite{2024ApJ...977....6G} could be resolved in ten resolution elements with a resolution of $10\;\mu$arcsec).

\subsection{Extreme Coronal Mass Ejections}
\label{sect:CMEs}
Recent X-ray studies of young, magnetically active stars strongly suggest that coronal mass ejections (CMEs) are a fundamental, yet as-yet unobserved, component of stellar high-energy activity. The Chandra Orion Ultradeep Project (COUP) revealed hundreds of powerful, long-duration X-ray flares in pre-main-sequence stars, with inferred magnetic loop sizes extending several stellar radii and, in some cases, plausibly connecting the stellar surface to the inner disk \citep{Favata2005,Getman2008}. These events provide the most compelling indirect evidence for stellar CMEs: by analogy with the Sun, such extreme flares should be accompanied by large-scale mass ejections. Building on this, \citet{Aarnio2012} used empirically calibrated solar flare--CME relationships to infer CME mass-loss rates of $10^{-12}$--$10^{-9}\,M_\odot\,\mathrm{yr^{-1}}$ for T Tauri stars, implying that CMEs may dominate early stellar mass and angular momentum loss and significantly impact disk evolution and planet formation environments. However, these conclusions remain indirect, as no stellar CME has yet been spatially resolved or unambiguously detected in X-rays, despite growing observational and theoretical efforts \citep{Drake2013,Osten2015,Argiroffi2019,Moschou2019,Leitzinger2014}.

A next-generation high-resolution X-ray imager with $\sim10\,\mu$as spatial resolution and 10--100$\times$ Chandra effective area would enable the first direct observational tests of stellar CME physics in nearby star-forming regions (e.g., Orion at $\sim$400 pc). At these distances, such angular resolution corresponds to spatial scales of a few stellar radii, sufficient to resolve flare sites, track expanding plasma structures, and distinguish confined coronal loops from escaping ejecta. Time-resolved imaging over flare durations (tens of ks) would allow detection of outward-moving, cooling plasma consistent with CME launch and propagation, while simultaneous spectroscopy (1--7 keV) would constrain densities, temperatures, and energetics. By observing statistically significant samples of young stars within a wide field of view, the mission could directly measure CME occurrence rates, energetics, and geometries, transforming current flare-based proxies into a physically grounded picture of stellar mass loss. This capability is essential for linking high-energy stellar activity to disk ionization, atmospheric erosion of young planets, and the early evolution of stellar angular momentum.

Since flares are stochastic, we need to observe a few young objects for a long time (Ms) or monitor many young objects simultaneously, requiring a FOV that can contain a significant fraction of the young stellar cluster.
For example, COUP used a 1~Ms observation of the central $\sim5^\prime$ of the ONC to detect $\sim$30 superflaring events with sufficient signal and time resolution to permit detailed modeling \citep{2005ApJS..160..469F}.

\subsection{Flares connecting star and proto-planetary disk}
\label{sect:flaredisk}
Young, cool stars are ubiquitous X-ray emitters. They are also still surrounded by their primordial disks and interact with them in different ways. XUV radiation from the disk ionizes the upper layers and drives disk chemistry. Disks are expected to be magnetically connected to the stellar magnetic field, and mass falls onto the star along those field lines (magnetically funneled accretion). Improved modeling has revealed how complex the disk structure is, with ionized outer layers, in- and out-flows, dead zones, and zombie zones \citep{2013ApJ...764...65M} with little or no magnetic field and possibly dust traps that concentrate grains \citep{2025Ap&SS.370..140P}. Understanding planet formation is not possible without understanding the disk structure, which determines where grains can coagulate, where planetesimals grow, and where they accrete gas atmospheres.

Yet, probing the magnetic structure of a star-disk system is currently limited to Zeeman-Doppler imaging of the stellar surface. The COUP data \citep{2005ApJS..160..469F,2008ApJ...688..437G} contain over a hundred superflares, which are bright and hot enough that the flare loop length would be sufficient to reach from the star to the surrounding disk. This is supported by the discovery of a wave pattern in the flare light curve in one case \citep{2018ApJ...856...51R} indicating that hot plasma is sloshing back and forth in the flare loop. However, the flare properties of stars with and without disks are indistinguishable \citep{2021ApJ...920..154G}, so spatially resolving flares is the only way to determine whether field lines connecting to the disk ignite flares and directly eject XUV radiation and energetic particles into the planet formation zone.

Because younger stars are more active, they are more likely to produce extreme observable events, and thus observations to resolve CMEs (section~\ref{sect:CMEs}) are best done on young stars. Whether flares occur in magnetic field lines that connect to the disk is a different physical question, but it looks at the same objects, so the observational requirements are identical to those in section \ref{sect:CMEs}.

%The brightest flares have tens of thousands of counts in Chandra and last 30-60~ks, sufficient signal to resolve them into of order 10 resolution elements.

\subsection{Protostars}
\label{sect:protostars}
Protostellar magnetic fields play a central role in star formation, mediating accretion and accelerating accretion-driven outflows. Observations indicate that most young protostellar objects accrete via their magnetospheres, where mass is transported from the inner disk to the surface of the (proto)star along magnetic field lines \citep{Hartmann_2016}.
%Mass is funneled from the inner protostellar disk to the surface of the protostar along the magnetic field lines \citep{Hartmann_2016}.
A combination of stellar and external fields amplified by gravitational collapse mediates outflows of collimated jets and wide-angle winds launched from the disks \citep{Ray_2021}. The outflows remove angular momentum, eject mass out of the system, determine the final mass of the forming star, and regulate the star formation rate on molecular cloud scales \citep{2021ApJ...912L..19P,2022MNRAS.515.4929G,2023ASPC..534..567P}.

The magnetic fields that mediate accretion and feedback are among the most fundamental and difficult-to-observe components of the star formation process.
The X-ray luminosity strongly correlates with the total magnetic flux from a star \citep[e.g.][]{Pevtsov_2003, Zhuleku_2020}, making X-ray emission the primary source of information on protostellar stellar magnetic fields.
While numerical simulations show that it is possible for dynamos powering magnetic fields in the central protostar to exist as early as the Class 0 phase \citep{Bhandare2020}, the observational evidence for this is scarce. X-ray detections of Class 0 protostars will provide the most substantial evidence that magnetic dynamos are active at such a young age.

Some Class I protostars are observed to be magnetically driven hard X-ray sources, often seen in flares \citep{Koyama_1996, Tsujimoto_2002, Tsuboi_2004}. X-ray flares can originate from reconnection events between stellar magnetic field lines or magnetically confined stellar plasma \citep{Feigelson_1999, Tsujimoto_2002}. In addition, X-ray emission has been detected 1-2$^{\prime\prime}$ from protostars along the axis of jets, which is suggested to arise from high-velocity (200-1000 km/s) bow shocks within the jet impacting slower-moving material \citep{Pravdo_2001, Tsujimoto_2004}. High angular resolution is needed to separate X-ray emission from outflow shocks from those due to stellar magnetic fields.

\begin{figure*}[h]
    \centering
    \includegraphics[width=0.95\linewidth]{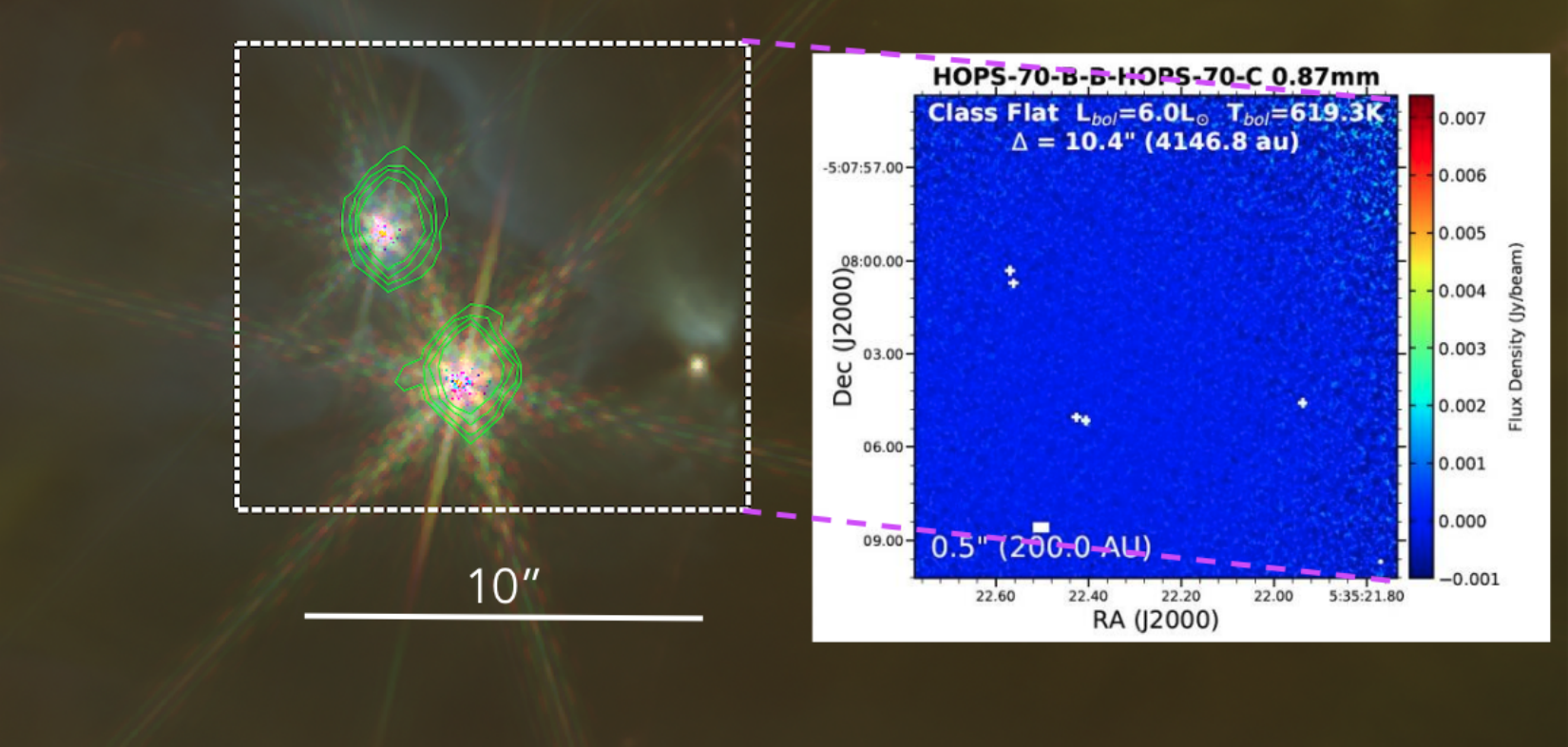}
    \caption{Left: A JWST NIRCAM image of HOPS-70.  The white box is about 14$^{\prime\prime}$ on a side. The green contours indicate Chandra flux. Right: An ALMA 0.87mm image of the boxed region. The two X-ray sources are further resolved into binaries separated by $\sim 0.25^{\prime\prime} $ (100 AU). https://planetstarformation.iaa.es/HOPS-70}
    \label{fig:HOPS-70}
\end{figure*}

The Herschel Orion Protostar Survey (HOPS) is a sample of 410 young stellar objects (YSOs) in the Orion molecular clouds, selected from Spitzer data. %Most objects have near-infrared photometry from 2MASS, mid- and far-infrared data from Spitzer and Herschel, and submillimeter photometry from APEX; thus, the SEDs cover 1.2 - 870 microns and are used to classify the sample into protostellar classes. Of the 410 YSOs, 330 have Spitzer and Herschel data and are mostly protostars; the remaining objects include likely extragalactic contaminants and faint YSOs. Using mid-IR spectral indices and bolometric temperatures, the sample of 330 YSOs is classified into 92 Class 0 protostars, 125 Class I protostars, 102 flat-spectrum sources, and 11 Class II pre-main-sequence stars.
Follow-up imaging with ALMA (Furlan et al. 2016) revealed that more than half of the HOPS sources are multiples, some with as many as 5 protostars. This is a much higher level of multiplicity than seen in the more evolved YSOs in Orion and may point to the importance of stellar interaction in the earliest stages of stellar evolution. This is out of reach of current instrumentation, even for nearby star-forming regions such as Orion. As shown in Figure~\ref{fig:HOPS-70}, each detected source is itself two sources. A HiReX mission could resolve and track the X-ray evolution of hundreds of such systems.

Unlike in the previous sections, these targets are too far away to resolve the flares themselves, here the goal is to identify and resolve the proto-stars and jets from each other. Relatively close-by stars on the main-sequence or in older star-forming regions are easily seen and separated in most wavelengths, but only X-rays and the far-IR or radio can penetrate the dust around deeply embedded star formation, so HiReX observations are needed to identify and separate sources in the first place.

\subsection{Jets from young stars}
\label{sect:jets}
Jets and outflows are ubiquitous phenomena in accreting astrophysical objects, from small brown dwarfs to supermassive black holes, and those outflows are believed to be launched by similar mechanisms. Young nearby stars are uniquely positioned to study this, as we can resolve details that remain a mystery even in supermassive systems due to their large distances. Learning how this universal phenomenon works can have a huge impact on many fields in astrophysics since these outflows carry away angular momentum, thus allowing accretion to proceed, while at the same time ejecting energy and turbulence in the surrounding medium, such as the star-forming cloud, the ISM, or the IGM. Outflows from young stars might be the driver that limits the efficiency of star formation in clouds \citep{Federrath_2014,Lebreuilly_2023}.

\begin{figure}
\centering
\includegraphics[width=\columnwidth]{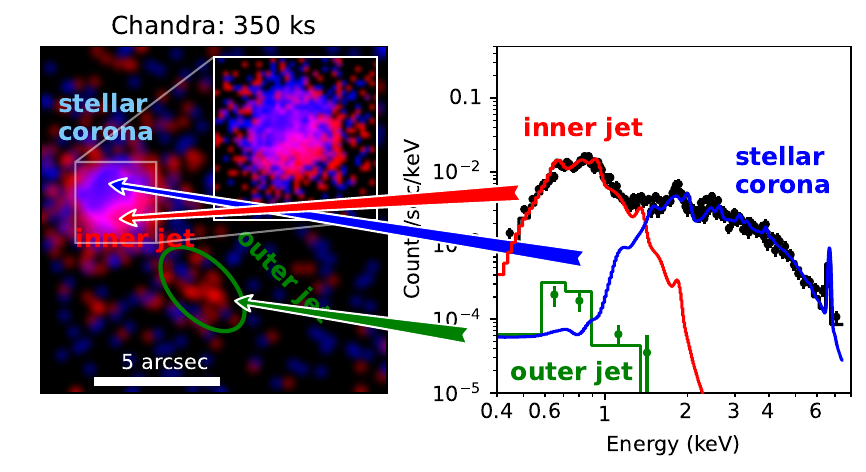}
\caption{Chandra observations of DG Tau. Chandra detects a spatial offset between the hard (blue) and soft (red) photons of order $0.3^{\prime\prime}$, but cannot resolve the components. In this particular case, the central star is heavily absorbed, leading to the hard (blue) spectrum, while the base of the jet is soft (red spectrum). Spatial resolution better than $0.1^{\prime\prime}$ would reveal the separate components and the shape of the inner jet.\label{fig:jets}}
\end{figure}

\begin{figure}
\centering
\includegraphics[width=\columnwidth]{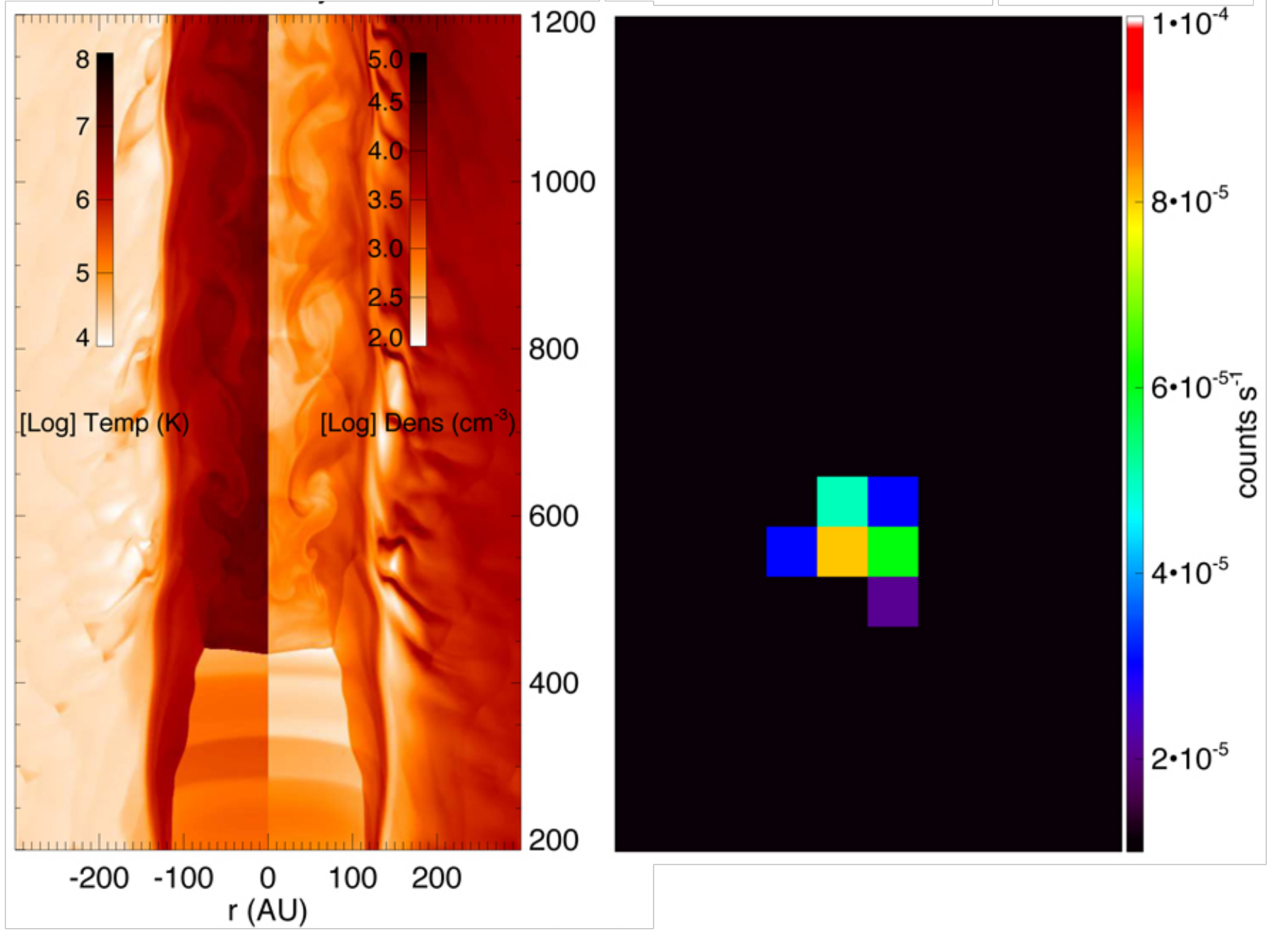}
\caption{Simulation of the HH~154 jet. The middle right panel shows the simulation, and the right panel shows synthetic X-ray data with $0.5^{\prime\prime}$ pixels. Clearly, that is insufficient to resolve even the most basic jet structures. Modified from a figure in \citet{2018A&A...615A.124U}. Reproduced with permission \copyright ESO.\label{fig:jetsim}}
\end{figure}

In young stars, most of the mass is launched magneto-centrifugally from the disk and moves at speeds between a few~km/s and 100-200~km/s. It is widely observed in molecular lines, the IR, and in the optical. Yet, jets often have an inner, faster-moving component that emits in X-rays (Figure~\ref{fig:jets}) and is presumably launched deep within the star's gravitational well, either from the star itself or in the star-disk interaction region \citep{2022hxga.book...57S}. Fast, shock-heated gas can be observed in X-rays. The outer components are seen with Chandra on the scale of arcseconds but have internal structure at much smaller scales (Figure~\ref{fig:jetsim}). Many jets also show emission very close to the star, which seems to be stationary \citep{2011ASPC..448..617G,2011A&A...530A.123S,2013A&A...552A.142G}. This component is barely distinguishable from the star itself in Chandra data (Figure~\ref{fig:jets}), and the current spatial resolution is insufficient to determine the cause of the X-ray emission. There are many questions that spatially resolved X-ray observations can answer, for example, is the material simply launched hot and cooling down over time? Or is it passing through a diamond shock \citep{2011ApJ...737...54B}, collimated by magnetic fields \citep{2018A&A...615A.124U}, or by the gas pressure of the outer wind components \citep{2014ApJ...795...51G}? We will not understand the energetics of the star-disk interaction, crucial for the disk lifetime, and the physics of star formation without resolving the jet launching zone. This requires a spatial resolution of the level of a few au in nearby star-forming regions (10~marcsec), with effective areas a few times larger than Chandra or XMM-Newton.

\section{Massive Stars}
\label{sect:massive}
Massive stars, i.e., spectral types OB and Wolf-Rayet (WR), are a population of stars that are intrinsic X-ray sources due to hypersonic shocks originating from their winds. Unlike lower-mass stars, massive stars do not form coronae. Instead, the significant UV-flux (blackbody temperatures of $T>10$\,kK) from the surface couples with the metal ions in their atmospheres to radiatively drive a wind to speeds of $v_\infty=1500 - 3000$\,km\,s$^{-1}$. The radiation driving also allows for a substantial amount of mass-loss from these stars, on the order of $10^{-8} - 10^{-5}\,M_\odot\,\mathrm{yr}^{-1}$, that alters both the surrounding interstellar environment and galactic feedback \citep{Krause2013}.

There are three broad categories for the types of wind shocks: (1) embedded wind shocks (EWS), (2) colliding wind shocks (CWS), or (3) magnetically-confined wind shocks (MCWS). In all cases, the shocks are believed to form at extended distances from the star, so the X-ray emission encodes information about the wind conditions. Yet, the limited resolution of current X-ray instruments is not sufficient to resolve the spatial extent or distribution of the X-ray emission.

\begin{figure}
    \centering
    \includegraphics[width=\columnwidth, trim=100 190 100 190]{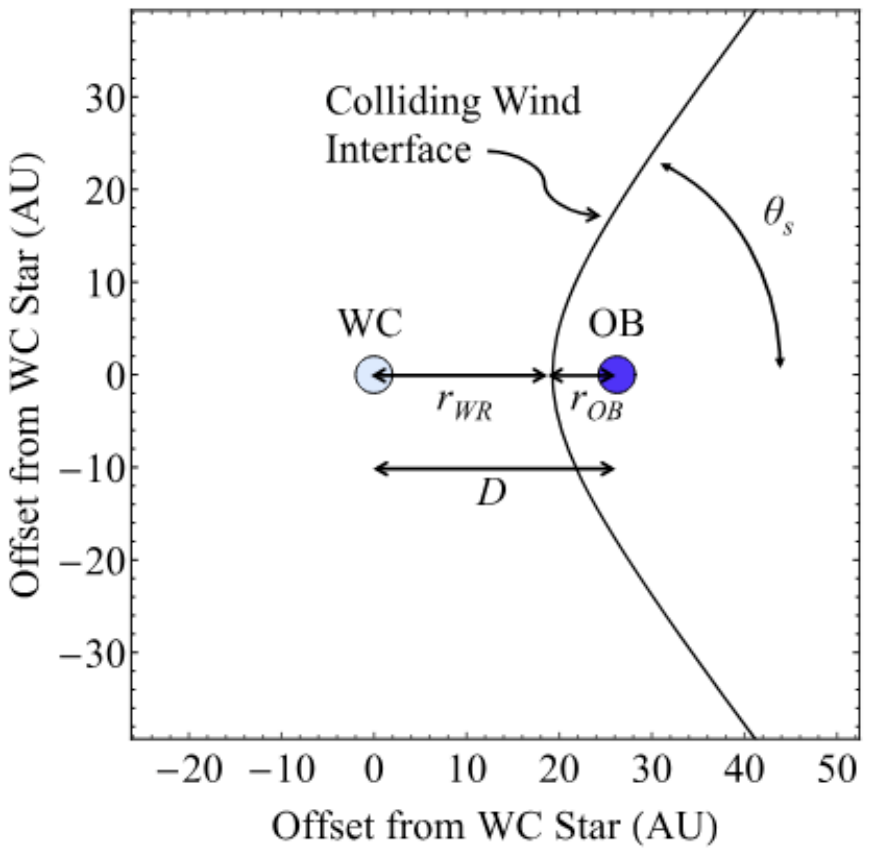}
    \caption{Example schematic of a WC+O colliding wind system. Reproduced with permission from the authors of \citet{Lau2020}.}
    \label{fig:CWSchematic}
\end{figure}

In the case of EWS stars, e.g., the prototype $\zeta$\,Puppis, the shocks are caused by intrinsic instabilities in the radiation driving \citep{Owocki2004}. These instabilities cause the wind to break up in clumps with varying accelerations that lead to shock speeds similar to those seen in supernova remnants. The shocks form approximately at $r=1.5 - 2\,R_*\approx10 - 30\,R_\odot$, meaning the resulting X-ray emission passes through the entire wind column. The X-rays can thus be used to constrain important parameters like the mass-loss rate \citep{Cohen2014,Gunderson2024}. The methods used to measure parameters, however, rely on a single assumption: spherical symmetry. The actual distribution of shocks in the wind is unknown, potentially leading to significant systematic uncertainties we may not be aware of. An imager with a spatial resolution $\leq0.1$\,mas would be capable of resolving the shock distribution and directly testing our theories of wind acceleration and clumping sources, as well as improving our determination of mass-loss rates.

CWS stars, e.g., the prototype WR\,140, typically have much larger structures due to shocks formed by the extended winds of the two stars. The resulting structure forms an extended surface whose shape depends on the difference in wind momenta. In the most dramatic case, the surface is referred to as a ``shock cone'' that wraps around the star with a weaker wind. For example, in the star WR\,112, an imager with spatial resolution $\leq0.1$\,mas would resolve not only the shock surface but also the individual EWS of the binary components.

Finally, MCWS represents the most complex form of massive-star X-ray emission. Typical MCWS systems have $B\geq 1$\,kG fields that constrain the wind to flow along the magnetic field lines, but here there are significant open questions about the structure of the field and wind distribution. In some systems, typically slowly rotating stars, the X-rays are believed to occur from the magnetic field forcing the wind to collide at the magnetic equator. This is believed to be the case in $\theta^1$\,Ori\,C, as a prototypical magnetic massive star \citep{udDoula2022}. Recent work has revealed an entirely different structure when the star is rapidly rotating. Here the field instead constrains the wind to a co-rotating, centrifugal disk that, at some distant radius, breaks open through a magnetic reconnection event \citep{Owocki2022}. Where the X-ray emission occurs in such an event remains unclear. Possible explanations include the reconnection event heating the trapped gas, like in a cool star flare, or gas streaming back up the field lines, shocking with the magnetic pole. Testing these models is possible with an imager with spatial resolution of $10\,\mu$\,as, as it will be able to resolve a low-radius MCWS and more distant reconnection events.

\section{Planetary Nebulae}
\label{sect:PN}
Planetary nebulae (PNe) mark the transition of low- to inter\-mediate-mass stars (1-8 M$_{\odot}$) from the AGB to white dwarfs. During this phase, the star donates at least half of its metal-enriched total mass to the interstellar medium (ISM). PNe thus strongly influence the chemical evolution of their host galaxies and subsequent generations of stars and planetary systems. It has long been understood that stars in this phase generate a tenuous, fast wind that overtakes the material that the star previously ejected in a slow, dense wind when it was on the AGB, thus revealing the hot stellar core. The faster wind sweeps up, shocks, and compresses the circumstellar material. However, advances in the last few decades have shown that this is an overly simplistic picture that cannot account for many observed aspects of PNe, including the high-energy phenomena that play a crucial role in PNe evolution.

Diffuse X-ray emission in PNe is understood to originate in the hot bubble formed by the fast winds of the central star, perhaps with contributions from the shocked collimated outflows that are now believed to be common in such sources \citep{2008NewA...13..563A, 2016ApJS..226...15F, 2017IAUS..323..104M}. However, in all observed PNe, the bubble is much cooler than expected \citep{2014ApJ...794...99F}. There is evidence to suggest that heat conduction plays a role in moderating its temperature \citep{2016ApJ...822L..19F}, but predictions from this model, such as the radial distribution of hot bubble emission \citep{2016ApJS..226...15F} and mixing of nebular and fast wind abundances \citep{2003PASP..115.1002M}, remain untested, as current X-ray telescopes do not have the sensitivity or spatial resolution needed. Ultra-high resolution X-ray imaging will provide the data necessary to test this. While distance determinations to PNe are often highly uncertain, it is reasonable to say that there are $>$10$^{3}$ PNe within about 5 kpc \citep{2024RMxAA..60..227H}, about 30\% of which can be expected to have diffuse X-ray emission \citep{2012AJ....144...58K}. An imager with spatial resolution $\le$ 0.1$^{\prime \prime}$ will easily resolve the radial distribution of hot bubble emission and detect differences in elemental abundances across many of them. For example, NGC 6543 has an X-ray flux = 8$\times 10^{-13}$ erg cm$^{-2}$ s$^{-1}$ \citep{2001ApJ...553L..69C} and is about 10$^{\prime \prime}\times20^{\prime \prime}$  on the sky. An imager with a 10$\times$ increase in resolution elements and 100$\times$ increase in effective area compared to Chandra will collect sufficient counts in 235 ks to differentiate between spectra extracted from across the nebula.

PNe can also exhibit point-source X-ray emission near the central star due to the central star's hot photosphere \citep{2000ApJS..129..295G}, which typically emits at energies $<$0.5 keV. However, about 30\% of central stars emit at higher energies \citep{2014ApJ...794...99F}, for reasons that remain unclear. It may arise from interactions with an unseen binary companion, wind shocks, fall-back of stellar wind material onto the central star, or magnetic reconnection events. The question of binarity is particularly important, as it has been suggested that the presence of a companion plays a major role in nonspherical mass loss and thus PNe morphology. An X-ray imager with 10 $\mu$arcsec spatial resolution will clearly detect potential PNe central star companions as close as 0.05 AU from their primary in systems within 500 pc.

\section{Cataclysmic Variables}
\label{sect:CV}
Novae, a subclass of cataclysmic variables (CV), are interacting binary systems in which a primary white dwarf (WD) star accretes hydrogen-rich material from a companion star. Once enough material accumulates on the surface of the WD, a thermonuclear runaway (TNR) ejects gas (enriched in CNO and other heavy elements, as well as various types of dust grains), drives shocks, and often produces a supersoft X-ray source while residual nuclear burning continues. As nearby, rapidly evolving TNR explosions, novae are valuable laboratories for studying accretion, mass ejection, shocks, and WD evolution in close binaries, and they may also provide insights into possible pathways toward Type Ia supernova progenitors \citep{Sala+2025}. Many novae also form dust within a few weeks to months of outburst, making them ideal nearby laboratories for investigating grain condensation in fast, irradiated, shock-rich outflows on a rapid time scale. Recent statistical work suggests that roughly 50–70\% of novae may form dust, with growing evidence that dust production is linked to radiative shocks in the ejecta \citep{Chong+2025}.

Current X-ray understanding of novae binaries has been built through observations with X-ray observatories such as Swift, Chandra, and XMM-Newton, combining high-cadence monitoring, deep pointed observations, and high-resolution spectroscopic studies. These studies have investigated the general behavior of the hard and supersoft X-ray phases across large samples and have provided detailed constraints on the physical and chemical conditions within the expanding ejecta. While these observatories have laid a strong foundation for the temporal and spectroscopic study of novae, further progress now requires Ultra-high resolution X-ray imaging. A first key science case is the direct imaging of nova shocks
\citep{Chomiuk+2021,Metzger+2014,Nelson+2021}. Hard X-ray emission in novae is generally associated with hot, shocked plasma produced when fast ejecta collide with slower, denser material or with pre-existing circumstellar gas \citep{Metzger+2014,Bose+2006,Nelson+2021}. Such shocks are also implicated in particle acceleration and, in many novae, GeV $\gamma$-ray production \citep{Abdo+2010,Metzger+2015,Chomiuk+2021}.
Yet without spatially resolved X-ray imaging, it is difficult to determine whether the X-rays arise from polar outflows, equatorial density enhancements, reverse shocks, clumps, or external interactions with
circumstellar material \citep{Nelson+2021,Chomiuk+2021,Bode+2006,Abdo+2010}. A second key science case is the connection between shocks, dust formation, and dust destruction in the novae ejecta. Dust formation in novae remains puzzling because grains condense rapidly in ejecta that are expanding, irradiated, and later exposed to supersoft X-rays from the central WD. Radiative shocks offer a promising solution: they compress gas into cool, dense post-shock regions that may shield molecules and grains from the intense radiation field and enable rapid grain growth \citep{Pandey+2024,Derdzinski+2017}. High-resolution X-ray imaging would test this scenario by locating hard X-ray shocks relative to infrared and submillimeter dust emission, identifying whether dust forms in post-shock cooling layers, dense clumps, or equatorial structures, and determining whether the supersoft X-ray source subsequently destroys, modifies, or bypasses the newly formed dust. Such observations are essential for distinguishing among competing dust-formation scenarios, identifying which ejection components produce or destroy dust, and measuring how shocks and irradiation shape the ejecta.

The importance of X-ray morphology is already demonstrated by Chandra observations of the old nova remnant GK Per, which show that the X-ray nebula is resolved, clumpy, and asymmetric rather than a simple spherical shell, highlighting that morphology itself carries crucial information about the shaping of the ejecta and its interaction with the surrounding medium \citep{Balman+2005}. An X-ray imager with spatial resolution $\leq 0.1^{\prime\prime}$ would extend such studies to a much larger sample of novae, separating the central WD and supersoft source from nearby shocked ejecta and resolving the clumpy and filamentary structures in which dust may form. Pushing to milliarcsecond or microarcsecond resolution would be transformative, enabling direct imaging of the shock and dust-forming regions during the first weeks to months of outburst. Such capability would connect the temporal and spectroscopic diagnostics established by current observatories with the missing spatial information needed to understand nova mass ejection, shock physics, dust formation, and WD evolution.

%------------------------------------------------
%\phantomsection
%\section*{Acknowledgments} % The \section*{} command stops section numbering

%\addcontentsline{toc}{section}{Acknowledgments} % Adds this section to the table of contents

%So long and thanks for all the fish.

%----------------------------------------------------------------------------------------
%	REFERENCE LIST
%----------------------------------------------------------------------------------------
\phantomsection
\include{journal_abbreviations}
\bibliographystyle{aasjournalv7}
\setlength\bibsep{0pt}
\bibliography{references}

%----------------------------------------------------------------------------------------

\end{document}

%% file: journal_abbreviations.tex
\newcommand{\sun}{$_{\odot}$\xspace}
\DeclareRobustCommand{\ion}[2]{%
\relax\ifmmode
\ifx\testbx\f@series
{\mathbf{#1\,\mathsc{#2}}}\else
{\mathrm{#1\,\mathsc{#2}}}\fi
\else\textup{#1\,{\mdseries\textsc{#2}}}%
\fi}

\def\farcm{\hbox{$.\mkern-4mu^\prime$}}
\def\farcs{\hbox{$.\!\!^{\prime\prime}$}}
\def\degr{\hbox{$^\circ$}}
\def\arcmin{\hbox{$^\prime$}}
\def\arcsec{\hbox{$^{\prime\prime}$}}

\def\na{New Astronomy}
\def\rmxaa{Revista Mexicana de Astronomía y Astrofísica}
\def\aj{AJ}%

          % Astronomical Journal

\def\araa{ARA\&A}%

          % Annual Review of Astron and Astrophys

\def\apj{ApJ}%

          % Astrophysical Journal

\def\apjl{ApJ}%

          % Astrophysical Journal, Letters

\def\apjs{ApJS}%

          % Astrophysical Journal, Supplement

\def\ao{Appl.~Opt.}%

          % Applied Optics

\def\apss{Ap\&SS}%

          % Astrophysics and Space Science

\def\aap{A\&A}%

          % Astronomy and Astrophysics

\def\aapr{A\&A~Rev.}%

          % Astronomy and Astrophysics Reviews

\def\aaps{A\&AS}%

          % Astronomy and Astrophysics, Supplement

\def\azh{AZh}%

          % Astronomicheskii Zhurnal

\def\baas{BAAS}%

          % Bulletin of the AAS

\def\jrasc{JRASC}%

          % Journal of the RAS of Canada

\def\memras{MmRAS}%

          % Memoirs of the RAS

\def\mnras{MNRAS}%

          % Monthly Notices of the RAS

\def\pra{Phys.~Rev.~A}%

          % Physical Review A: General Physics

\def\prb{Phys.~Rev.~B}%

          % Physical Review B: Solid State

\def\prc{Phys.~Rev.~C}%

          % Physical Review C

\def\prd{Phys.~Rev.~D}%

          % Physical Review D

\def\pre{Phys.~Rev.~E}%

          % Physical Review E

\def\prl{Phys.~Rev.~Lett.}%

          % Physical Review Letters

\def\pasp{PASP}%

          % Publications of the ASP

\def\pasj{PASJ}%

          % Publications of the ASJ

\def\qjras{QJRAS}%

          % Quarterly Journal of the RAS

\def\skytel{S\&T}%

          % Sky and Telescope

\def\solphys{Sol.~Phys.}%

          % Solar Physics

\def\sovast{Soviet~Ast.}%

          % Soviet Astronomy

\def\ssr{Space~Sci.~Rev.}%

          % Space Science Reviews

\def\zap{ZAp}%

          % Zeitschrift fuer Astrophysik

\def\nat{Nature}%

          % Nature

\def\iaucirc{IAU~Circ.}%

          % IAU Cirulars

\def\aplett{Astrophys.~Lett.}%

          % Astrophysics Letters

\def\apspr{Astrophys.~Space~Phys.~Res.}%

          % Astrophysics Space Physics Research

\def\bain{Bull.~Astron.~Inst.~Netherlands}%

          % Bulletin Astronomical Institute of the Netherlands

\def\fcp{Fund.~Cosmic~Phys.}%

          % Fundamental Cosmic Physics

\def\gca{Geochim.~Cosmochim.~Acta}%

          % Geochimica Cosmochimica Acta

\def\grl{Geophys.~Res.~Lett.}%

          % Geophysics Research Letters

\def\jcp{J.~Chem.~Phys.}%

          % Journal of Chemical Physics

\def\jgr{J.~Geophys.~Res.}%

          % Journal of Geophysics Research

\def\jqsrt{J.~Quant.~Spec.~Radiat.~Transf.}%

          % Journal of Quantitiative Spectroscopy and Radiative Trasfer

\def\memsai{Mem.~Soc.~Astron.~Italiana}%

          % Mem. Societa Astronomica Italiana

\def\nphysa{Nucl.~Phys.~A}%

          % Nuclear Physics A

\def\physrep{Phys.~Rep.}%

          % Physics Reports

\def\physscr{Phys.~Scr}%

          % Physica Scripta

\def\planss{Planet.~Space~Sci.}%

          % Planetary Space Science

\def\procspie{Proc.~SPIE}%

          % Proceedings of the SPIE

\def\nar{New Astronomy Reviews}

\let\astap=\aap

\let\apjlett=\apjl

\let\apjsupp=\apjs

\let\applopt=\ao

\uchyph=0